\documentstyle[aas2pp4,flushrt]{article}
\input epsf.sty
\let\simgt=\gtrsim
\newcommand{\gdm}{{g}}
\newcommand{\eff}{{\rm eff}} 
\newcommand{\vis}{{\rm vis}} 
\newcommand{\hdm}{{\rm hdm}} 
\newcommand{\scpot}{V}

\begin{document}

\title{Structure Formation with Generalized Dark Matter}

\author{Wayne Hu\footnote{Alfred P. Sloan Fellow}}
\affil{Institute for Advanced Study, Princeton, NJ 08540}

\begin{abstract}
The next generation of cosmic microwave background (CMB) experiments,
galaxy surveys, and high-redshift observations
can potentially determine the nature of
the dark matter observationally.  With this in mind, we 
introduce a phenomenological model for a generalized
dark matter (GDM) component and discuss its effect on large-scale 
structure and CMB anisotropies.  Specifying the gravitational 
influence of the otherwise non-interacting GDM requires not merely
a model for its equation of state but one for its full stress tensor.
From consideration of symmetries, conservation laws, and gauge invariance, 
we construct a simple but powerful 3 component parameterization of 
these stresses that exposes the new phenomena
produced by GDM.  Limiting cases include: a particle component
(e.g.\ WIMPS, radiation or massive neutrinos), a cosmological
constant, and a scalar field component.
Intermediate cases illustrate how the  
clustering properties of the dark matter can be specified independently of 
its equation of state.  This freedom allows one to alter the
amplitude and features in the matter power 
spectrum relative to those of the CMB anisotropies while leaving
the background cosmology fixed.
Conversely, observational constraints on such phenomena can
help determine the nature of the dark matter.  
\end{abstract}

\section{Introduction}
\label{sec:intro}

Upcoming cosmic microwave background
(CMB) missions, galaxy redshift surveys, and high-redshift 
observations will produce such 
a wealth of high-quality data that even the extended cold dark matter
(CDM) model with 11 free parameters (e.g.\ \cite{Jun96}\ 1996) 
may fail to fit them. One must face the very real possibility that 
none of our current {\it ab initio} models will survive the upcoming
confrontation with the data.  

How might one generalize the CDM model?
The cornerstone of all modern cosmologies
is, and will likely remain, gravitational instability in a
world-model that is homogeneous and isotropic on the large scale
(e.g.\ \cite{Pee91}\ 1991).  Models for the dark matter sector, on the
other hand, are presently limited by the number of candidates
that are considered well-motivated from the particle physics 
standpoint, e.g.\ WIMPS (CDM), massive neutrinos, scalar fields 
(see \cite{Cob97}\ 1996; 
\cite{Fer97}\ 1997; 
\cite{Cal98}\ 1998, 
for recent assessments) 
and topological defects (\cite{Pen97}\ 1997).  If none of these candidates
survive the confrontation with high-precision cosmological measurements, 
we will be forced to solve the inverse problem:
can one determine the
nature of the dark matter and reconstruct the model for structure
formation directly from the observations? 

Explorations of the dark matter sector have been undertaken recently
by \cite{TW97} (1997) and \cite{Cal98} (1998) 
who considered dark matter with an arbitrary equation of state that possesses
no fluctuations and
scalar-field type fluctuations respectively.  The former case does not 
present a complete theory of structure formation as it can only apply
below the horizon at any given time for adiabatic models. 
The latter case provides an interesting example of an exotic dark matter component but
does not exhaust the possibilities for its gravitational properties.  For example, 
a hot dark matter component has an equation of state like matter today
but clusters only up to a finite scale.  This scale 
is substantially below the current horizon scale, where it would be 
for the analogous scalar field model (\cite{Fer97}\ 1997).

Starting from the general principles of symmetry, energy-momentum conservation 
and gauge invariance, we build 
in \S \ref{sec:properties} 
a phenomenological parameterization of the dark
matter 
that includes all of the currently 
popular dark matter theories as special cases.    The main result is the link
established between 
the clustering properties of the dark matter and the model for its
underlying stresses.   Importantly, these properties are not determined by the background 
equation of state in the general case.  Through their effect on the growth rate of
perturbations, discussed in \S\ref{sec:growth}, these clustering properties manifest themselves themselves
as independent 
features in the power spectra of large-scale structure and CMB anisotropies as discussed 
in \S \ref{sec:lss} and \S \ref{sec:CMB}.  We summarize the main results in \S \ref{sec:conclusions} and present a short Appendix that highlights the
scalar field case.

\section{Dark Matter Properties}
\label{sec:properties}

\subsection{General Principles}

The gravitational influence of an arbitrary dark matter 
component is controlled by its stress-energy tensor
$T_{\mu\nu}({\bf x},\eta)$, where $\eta = \int dt/a$
is the conformal time with $a(t)$ as the scale factor normalized
to unity today.  In general, the symmetric 4-tensor
$T_{\mu\nu}$ has 10 components which can be divided into
4 classes: the energy density $\rho_\gdm$ (1), 
the isotropic stress or pressure $p_\gdm$ (1),
the momentum density $(\rho_\gdm+p_\gdm)v_\gdm^i$ (3),
and the anisotropic stress
$p_\gdm \pi_\gdm^{ij}$ (5).  The 5 components of the anisotropic 
stress can be further separated by their transformation properties under rotations
into a scalar component (1), vector components (2), and tensor 
components (2).  Since only scalar components exhibit gravitational
instability, we hereafter neglect the vector and tensor 
contributions.\footnote{Vector and tensor contributions do however
affect CMB anisotropies and can act as additional degrees of
freedom when normalizing large-scale structure to CMB measurements (e.g.\ \cite{Cal98b}\ 1998).}
Energy-momentum conservation 
$T_{\mu\nu}^{\hphantom{\mu\nu};\nu}=0$ introduces 4 constraints, leaving only
two independent parameters for the dark matter.  One can choose
these to be the pressure $p_\gdm$ and scalar anisotropic 
stress amplitude $\pi_\gdm$ [see \cite{Bar80} 1980, Eq. (2.18)] without loss of generality.  

The isotropy of the background implies that the anisotropic stress
and momentum density can only be present as a perturbation.   
The conservation laws then reduce to a single relation,
\begin{equation}
{\dot\rho_\gdm \over \rho_\gdm} = -3(1+w_\gdm){\dot a \over a}\,,
\label{eqn:density}
\end{equation}
where overdots represent conformal time derivatives and 
$w_\gdm = p_\gdm/\rho_\gdm$.  Likewise the Einstein equations
$G_{\mu\nu}=8\pi GT_{\mu\nu}$ reduce to
\begin{equation}
\left( {\dot a \over a} \right)^2 = {8\pi G \over 3} a^2 \sum_i \rho_i - K\,,
\end{equation}
where the sum is over the density contributions of all matter species and the
background curvature is $K= -H_0 (1 - \Omega_{\rm tot})$ with 
$\Omega_{\rm tot}=\sum_i \Omega_i$. 
As usual the expansion rate today is given by the
Hubble constant $H_0 \equiv (\dot a/a)_{a=1} =100 h\,$km s$^{-1}$ Mpc$^{-1}$,
to which each species contributes according to its fraction of the critical
density $\Omega_i=8\pi G\rho_i/3H_0^2$.

The background evolution is thus completely specified by the
equation of state $w_\gdm(a)$.   
This is {\it not} true of the perturbations, 
where we are left with the freedom to specify $\delta p_\gdm$ 
and $\pi_\gdm$.  
It is convenient to separate out the non-adiabatic
stress or entropy contribution
\begin{equation}
p_\gdm \Gamma_\gdm =
\delta p_\gdm - c_\gdm^2 \delta \rho_\gdm \,,
\end{equation}
where the adiabatic sound speed is
\begin{equation}
c_\gdm^2 = {\dot p_\gdm \over \dot \rho_\gdm}\, 
         = w_\gdm - {1 \over 3} {\dot w_\gdm \over 1 + w_\gdm}
		\left( {\dot a \over a} \right)^{-1} .
\label{eqn:sound}
\end{equation}
Therefore, $p_\gdm = w_\gdm \rho_\gdm$ does {\it not} imply
$\delta p_\gdm = w_\gdm \delta\rho_\gdm$ (due to temporal or spatial 
variations in $w_\gdm$)
and furthermore if $\Gamma_\gdm \ne 0$,
the function $w_\gdm(a)$ does not completely specify the pressure fluctuation.

In the following, we adopt the notation of \cite{OTamm} (1998).
For brevity, we present only the aspects of perturbation theory that are
altered
by the presence of GDM.  
Energy-momentum conservation yields the continuity equation
for the density fluctuation $\delta_\gdm\equiv \delta\rho_\gdm/\rho_\gdm$
\begin{equation}
\left( { \delta_\gdm \over {1 + w_\gdm}}\right)
\dot{\vphantom{\Big)}}
= -(k v_\gdm + 3 \dot h_\delta)  - 3 {\dot a \over a} {w_\gdm \over
	1+w_\gdm} \Gamma_\gdm \,,
\label{eqn:continuity}
\end{equation}
and the Euler equation
\begin{eqnarray}
{\dot v_\gdm} &=& - {\dot a \over a}(1-3c_\gdm^2) v_\gdm
+ {c_\gdm^2 \over 1 + w_\gdm} k \delta_\gdm  
+ {w_\gdm \over 1+w_\gdm} k\Gamma_\gdm 
\nonumber\\
&& 
-{2 \over 3} {w_\gdm \over 1+w_\gdm} (1-3K/k^2)k\pi_\gdm
+ k h_v \,.
\label{eqn:euler}
\end{eqnarray}
The metric sources $h_\delta$ and $h_v$ depend on the choice of
gauge and are
\begin{eqnarray}
h_\delta &= &\cases{ h_L & Synchronous,\cr
		   \Phi_{\hphantom{XX}}& Newtonian,\cr } \nonumber\\
h_v      &=& \cases{ 0 & Synchronous, \cr
		   \Psi_{\hphantom{XX}}& Newtonian, \cr} 
\label{eqn:metric}
\end{eqnarray}
Note that $h_L = h/6$ in the notation of \cite{Ma95} (1995).
Seed perturbations (e.g.\ defects) whose total contribution
is first order in the perturbations can also be modeled in this 
manner by rewriting equations (\ref{eqn:density}), (\ref{eqn:continuity}) 
and (\ref{eqn:euler}) in terms of $\delta \rho_\gdm$ and
$(p_\gdm + \rho_\gdm) v_\gdm$ instead of the relative perturbations
(see eq.~[\ref{eqn:gdmscalarfields}]).

\subsection{Stress Model}

Up to this point, we have made no assumptions whatsoever about the
nature of the dark matter since $w_\gdm$, $\Gamma_\gdm$ and $\pi_\gdm$
have been left as free functions.   We now need to parameterize
these functions.  Dark matter with $w_\gdm < 0$ is favored
by current observational constraints,  such as the combination of
the ages of globular clusters and 
the high Hubble constant measurements (\cite{Chi97}\ 1997; \cite{Cal98}\ 1998) as well as supernova luminosity distance measures
(\cite{Per98} 1998;\cite{Rie98} 1998). 
If $w_\gdm < 0$ and is slowly-varying
compared with the expansion rate $(\dot a /a)$
such that $c_\gdm^2 < 0$,  the adiabatic pressure fluctuation produces
accelerated collapse rather than support for the density perturbation.
In a GDM-dominated universe, perturbations would rapidly go nonlinear
once the sound horizon has been crossed  $| k\int c_\gdm d\eta| > 1$.
This situation is unacceptable for a model of structure formation.

In this $w_\gdm<0$ regime, it is interesting to consider whether 
non-adiabatic pressure can act to stabilize the
perturbation.  This requires a relation of the type $w_\gdm\Gamma_\gdm
\propto \delta_\gdm$. 
One is not however allowed complete freedom in 
establishing this relation.  Adiabatic pressure, density,
and velocity perturbations are  
gauge dependent whereas non-adiabatic pressure
perturbations are not. Therefore, stabilization in one frame of
reference does not equate to stabilization in another.  
One should avoid having the properties of the GDM depend 
on an arbitrary choice of frame and hence unphysically on 
the perturbations in the other species.  This requirement 
can be achieved by
defining the relation in the rest frame of the
GDM where 
$T^0_{\hphantom{0} i}=0$, 
\begin{equation}
w_\gdm \Gamma_\gdm = 
(c_\eff^2 - c_\gdm^2) 
\delta_\gdm^{\rm (rest)}\,.
\label{eqn:gammaansatz}
\end{equation}
We further assume the 
effective sound speed $c_\eff^2$ is only a function of time. 
If $c_\eff^2 > 0$, pressure support is obtained.  
The gauge transformation into an arbitrary frame gives 
\begin{equation}
\delta_\gdm^{(\rm rest)} = \delta_\gdm + 3{\dot a \over a}(1+ w_\gdm)
(v_\gdm - B)/k \,,
\end{equation}
yielding a manifestly gauge-invariant form for the non-adiabatic
stress (see \cite{Bar80}\ 1980; \cite{Kod84}\ 1984).
Here $B$ represents the time-space component of metric fluctuations
and vanishes in both the synchronous and Newtonian gauges.
The Euler equation can then be rewritten as
\begin{eqnarray}
\dot v_\gdm &=&- {\dot a \over a} v_\gdm +
{c_\eff^2 \over 1+w_\gdm} k\delta_\gdm^{(\rm rest)} \nonumber\\
&& - {2 \over 3}{w_\gdm \over 1+w_\gdm} k (1-3K/k^2)\pi_\gdm + k h_v \,.
\label{eqn:euleransatz}
\end{eqnarray}
Thus $c_\eff^2$ may be thought of as a rest-frame sound speed. 
By inspection of equations (\ref{eqn:euleransatz}) and (\ref{eqn:continuity}),
we determine that the critical scale for stabilization is the effective sound
horizon, 
\begin{equation}
s_\eff = \int d\eta c_\eff\,.
\end{equation}
This assumes $w_\gdm$ is not varying at a rate
much greater than the expansion rate (see Appendix). 

The anisotropic stress can also affect the density perturbations.
A familiar example is that of fluid, where it represents
viscosity and damps density perturbations.  More generally,
the anisotropic stress component is the amplitude of a 3-tensor that
is linear in the perturbation. 
A natural choice for its source 
is $k v_\gdm$, the amplitude of the velocity shear tensor $\partial^i v^j_\gdm$.
However it must also be gauge invariant and generated by the 
corresponding shear term in the metric fluctuation $H_T$.
The relationship between velocity/metric shear and anisotropic stress
may be parameterized with a ``viscosity parameter'' $c_\vis^2$
\begin{equation}
w_\gdm \left( \dot\pi_\gdm + 3{\dot a \over a}\pi_\gdm \right) = 
4 c_\vis^2  (k v_\gdm - \dot H_T)\, ,
\label{eqn:pi}
\end{equation}
where in the Newtonian gauge $H_T=0$ and in the 
synchronous gauge $H_T=h_T (= -h/2 - 3\eta$, \cite{Ma95}\ 1995).
The specific form of this equation is designed to recover
the free streaming equations of motion for radiation
with an approximate closing of the angular
moment hierarchy at the quadrupole (\cite{Hu95}\ 1995).
The physical interpretation of equation (\ref{eqn:pi}) is that
the anisotropic stress will act to damp out velocity 
fluctuations on shear-free frames ($H_T=0$) 
if $c_\vis^2 > 0$.  We call $s_\vis = \int c_\vis d\eta$ the 
viscous scale.

We shall see in the following sections that this parameterization
captures many of the essential features of GDM
as it corresponds to a means of altering its clustering
properties.  In the limit that
$(w_\gdm,c_\eff^2,c_\vis^2) \rightarrow (w_\gdm,1,0)$
scalar-field dark matter is recovered exactly (see Appendix).  
CDM is similarly recovered with $(0,0,0)$, radiation can be modeled as $(1/3,1/3,1/3)$
and hot dark matter (HDM) or warm dark matter in a similar matter described in 
\S \ref{sec:MDM}.  

Cases that this ansatz does {\it not} cover
involve mainly models in which stress fluctuations are not
derived from density and velocity perturbations and so act as
external sources for the perturbations.  This situation occurs
in the case of non-linear
seed perturbations (see \cite{HSW} 1997 for a parallel treatment).  Note that
in such models both vector and tensor stresses must also be modeled
to yield a complete theory for structure 
formation (see 
\cite{TAMM}\ 1997b;
\cite{Tur97}\ 1997).

\begin{figure}[hbt]
\centerline{ \epsfxsize = 3.25truein \epsffile{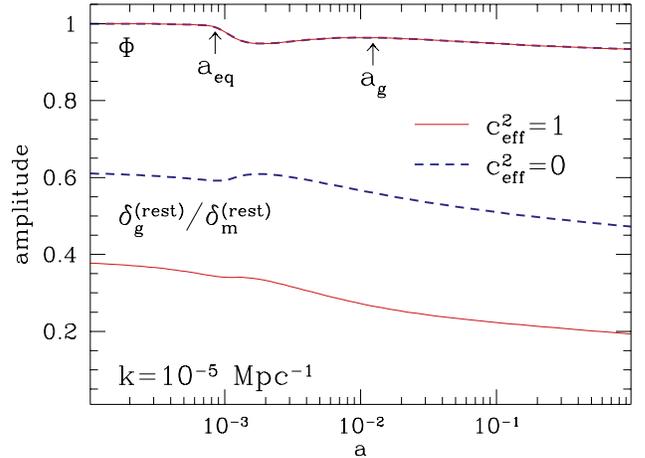}}
\caption{Large scale perturbation evolution ($k s_\eff \gg 1$) with GDM of $w_\gdm = -1/6$,
$c_\eff^2 = 1$ (solid, scalar fields) and $0$ (dashed, stress-gradient free).
The Newtonian curvature $\Phi$ is independent of $c_\eff^2$ and varies
only when the background equation of state changes at $a_{\rm eq}$ and $a_\gdm$.
However, the ratio of density perturbations in the GDM and matter 
depends on $c_\eff^2$.  The cosmological parameters here are $\Omega_{\rm tot}=1$,
$\Omega_\gdm=0.9$, $\Omega_b h^2=0.0125$, $h=0.7$.}
\label{fig:large}
\end{figure}

Equations (\ref{eqn:continuity}), (\ref{eqn:gammaansatz}),
and (\ref{eqn:euleransatz}) can now be introduced into
any of the standard codes that solve the 
Einstein-Boltzmann equations.  We employ the code of \cite{Hu95}\ (1995)
for the examples below.  
We furthermore show 
adiabatic models for illustrative purposes 
but of course isocurvature models can be similarly obtained through a change
in the initial conditions. 

\section{Perturbation Growth}
\label{sec:growth}
By introducing a means by which fluctuations in the GDM are stabilized,
we change the growth rate of fluctuations in the baryons and
any CDM that may be present.  The clustering properties of the
GDM thus have consequences for both large-scale structure
and CMB anisotropies as we shall see in \S \ref{sec:lss} and
\ref{sec:CMB}.  Here we summarize results for the growth rate of
perturbations proven in \cite{Hu98} (1998). 

\subsection{Above the Sound Horizon}
\label{sec:large}

Above the stabilization
scale of all species and in the absence of background
curvature $(K=0)$, perturbations evolve so as to keep the
Newtonian curvature $\Phi$ and potential $\Psi$ constant 
\begin{eqnarray}
(k^2 - 3K)\Phi & = & 4\pi G a^2 \sum_i \rho_i \delta_i^{(\rm rest)}
	\,,\nonumber\\
k^2(\Psi + \Phi)& = & -8\pi G a^2\sum_i p_i \pi_i \,,
\label{eqn:Poisson}
\end{eqnarray}
except during periods when the dominant equation of state $w$ changes.
Here the potentials vary mildly such that (\cite{Bar80}\ 1980)
\begin{equation}  
\Phi - {2 \over 3} {1 \over 1 + w } \Psi = {\rm const.}
\label{eqn:phiw}
\end{equation}
For reference, dominance of GDM relative to matter occurs at
\begin{eqnarray}
a_\gdm = \left( {\Omega_\gdm \over \Omega_m} \right)^{1/3w_\gdm} . 
\end{eqnarray}
The result is that during periods when the equation of state is
slowly-varying, the total density fluctuation $\delta^{\rm (rest)} 
= \sum \delta_i^{\rm (rest)} \rho_i /
\sum \rho_i$ grows as 
\begin{equation}
\delta^{(\rm rest)} \propto a^{1+3w} \,.
\label{eqn:grow}
\end{equation}
We display an example of this large-scale evolution in Fig.~\ref{fig:large}.

Since CMB anisotropies are only dependent on the time evolution of
the potentials, the effect of GDM here is very weak. The exception is
the $w_\gdm \rightarrow -1$ limit, where the potentials decay to zero and
the cosmological constant case is recovered.   Note however that
a curvature term in the background and a dark matter component with
$w_\gdm=-1/3$ give identical contributions to the expansion rate
but are not similar in their contribution to large-angle CMB
anisotropies.  

Finally, the division of density fluctuations between
the GDM and the matter components, i.e. $\delta_\gdm/\delta_m$,
depends on the form of the GDM stress.  This is because
of the appearance of the non-adiabatic stress term in the continuity
equation (\ref{eqn:continuity}) which relates the density and metric
fluctuations.   As $c_\eff^2$ increases, 
$\delta_\gdm/\delta_m$  decreases.  This affects the amplitude of
the matter power spectrum as we shall see (\S \ref{sec:soundhorizon}).

\begin{figure}[hbt]
\centerline{ \epsfxsize = 3.25truein \epsffile{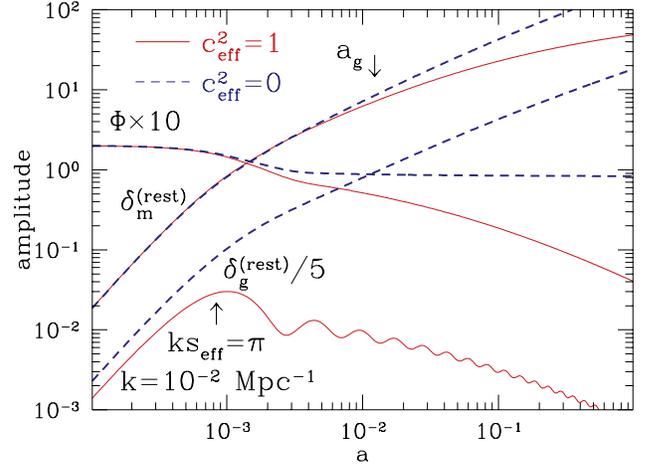}}
\caption{Small-scale perturbation evolution for the GDM and the matter density
perturbations in the same models as Fig.~\protect{\ref{fig:large}}
with $c_\eff^2=1$ (dashed lines) and $c_\eff^2=0$ (solid lines).  GDM perturbations
stabilize once $k s_{\rm eff} > \pi$ and their relative absence 
($\delta_g^{(\rm rest)}/\delta_m^{(\rm rest)} \ll 1$)
then slow the growth of matter perturbations once the expansion is 
also GDM dominated $a>a_g$ leaving the potential $\Phi$ to decay.}
\label{fig:small}
\end{figure}

\subsection{Below the Sound Horizon}
\label{sec:small}

Below the effective sound horizon of the
GDM, its perturbations stabilize.  If GDM dominates the 
energy density, $\delta_g^{(\rm rest)}$ oscillates with a decaying amplitude
of
\begin{equation}
A \propto a^{(-1+3w_\gdm)/2} c_\eff^{-1/2} \,,
\end{equation}
and rapidly becomes a smooth density component compared with
the fluctuations in the other species.  
We display a case where there is also a non-negligible CDM component
in Fig.~\ref{fig:small}.  
Once GDM dominates the energy density of the universe at $a_\gdm$, 
then the smoothing of the GDM-component will also slow or halt the
growth in the matter species.  In particular, the growth 
will slow to a halt if $w_\gdm < 0$ and scale as $a^p$ with 
$4p=\sqrt{1+24\Omega_m/\Omega_{\rm tot}}-1$ if $w_\gdm =0$.  Analytic
solutions for how this process occurs are given in \cite{Hu98} (1998). 

The net effect is
that if there is enough CDM to ensure that the universe was matter dominated
sometime in the past, 
perturbation growth is suppressed by a scale-independent 
factor below 
the effective sound horizon $s_\eff$ at GDM-domination.  This suppression
decreases until it disappears above $s_\eff$ today.   For the CMB, the suppression
of growth in the density perturbations causes the potentials to decay to
zero, leading to a contribution potentially $36$ times larger than the Sachs-Wolfe
effect in the anisotropy power spectrum.
We shall see in \S \ref{sec:ISW} why this limiting value is never reached 
in practice.

On the other hand, if matter never dominated the expansion as is the
case if CDM is absent, 
even more dramatic effects occur.  
In this case, even for $c_\eff^2=0$, growth is highly suppressed
on small scales due to an extended period of radiation domination
(assuming $w_\gdm < 0$). The controlling scale is therefore the horizon at
GDM-radiation equality
\begin{equation}
a_{\gdm\rm eq} = 
  \left( {\Omega_r \over \Omega_\gdm} \right)^{1 \over 1 - 3w_\gdm},
\label{eqn:equality}
\end{equation}
i.e.
\begin{equation}
k_{\gdm\rm eq} = \left( {\dot a \over a} \right)_{a = a_{\gdm \rm eq}}=
		\sqrt{2 \over\Omega_r} \Omega_\gdm H_0
		({ \Omega_r \over \Omega_\gdm } )^{-3 w_\gdm \over 1- 3 w_\gdm}.
\label{eqn:kequality}
\end{equation}
Above this scale, perturbations grow as in equation~(\ref{eqn:grow});
below this scale, perturbations only experience significant growth after
GDM-domination.
Because GDM-domination for $w_\gdm < 0$ is delayed compared with
an equivalent $w_\gdm =0$ universe, the critical scale is larger in 
such single component GDM models than in the CDM case.

\begin{figure}[hbt]
\centerline{ \epsfxsize = 3.25truein \epsffile{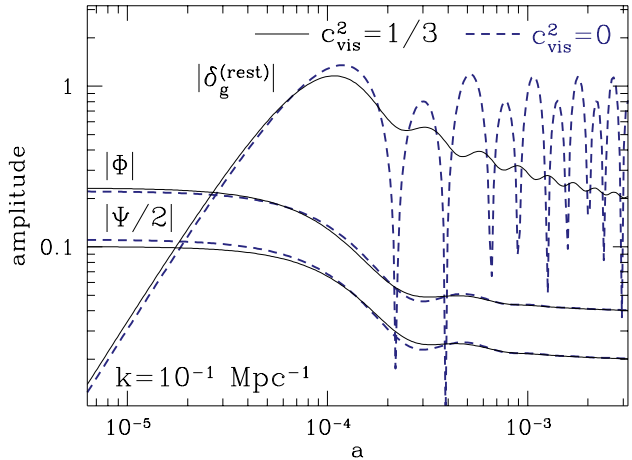}}
\caption{Viscous effects with GDM $w_\gdm=1/3$ replacing the three species of
massless neutrinos in the sCDM model ($\Omega_{\rm tot}=1 \approx \Omega_m$, $h=0.5$, $\Omega_b h^2=0.0125$).
With the viscosity parameter set to mimic radiation $c_\vis^2=1/3$ (solid lines) the
perturbations in the GDM decay whereas with $c_\vis^2=0$, they do not.  
This distinction has a negligible effect on the behavior of the 
potentials $\Phi$ and $\Psi$ well after sound horizon crossing}
\label{fig:timepi}
\end{figure}

\subsection{Viscosity}
\label{sec:viscosity}

Finally the viscous stresses of equation~(\ref{eqn:pi})
 can dissipate fluctuations in the GDM.  
In Fig.~\ref{fig:timepi}, we show an example with $w_\gdm=1/3$ GDM replacing
the three massless neutrinos in an otherwise standard CDM universe (sCDM, 
$\Omega_{\rm tot}=1 \approx \Omega_m$, $h=0.5$, $\Omega_b h^2=0.0125$).  
The real neutrino background radiation contains an anisotropic stress due to
the quadrupole moment of its temperature distribution.  Modeling the neutrinos 
as GDM allows us to explore the consequences of its anisotropy by varying
$c_\vis^2$ and also illustrates the phenomenological manifestations of the
viscosity parameter.

In the absence of anisotropic stresses (dashed lines, $c_\vis^2=0$),
perturbations
in the GDM oscillate.  Changing $c_\vis^2$ to $1/3$ to approximate the
radiative viscosity of the real neutrinos, the density perturbations damp once
$ks_\vis \simgt \pi$.  Note however, its effect on the gravitational potentials
$\Phi$ and $\Psi$ well after sound horizon crossing when large-scale structure is
observed is negligible.  This is because the pressure fluctuations are sufficiently effective
to make the GDM perturbations smooth in comparison to the growing species 
(see Fig.~\ref{fig:small}).  
The extra smoothing due to viscous damping affects perturbations little.  

The anisotropic stress does change the behavior of the potentials
at early times because it enters directly in to the Poisson
equations (\ref{eqn:Poisson}).
We shall see in \S \ref{sec:special} that correspondingly viscous effects 
are more important for the CMB than large-scale structure so long as
$|c_\eff^2| > |c_\vis^2|$.  

\section{Large-Scale Structure}
\label{sec:lss}

The large-scale structure of the
universe depends on the detailed properties of the GDM.  
The main result is that the clustering scale 
becomes independent of the
equation of state of the dark matter.  By changing the 
growth rate of perturbations,
the clustering properties change 
the amplitude of and features in the matter power spectrum.  Here
we present concrete examples of this process that include
scalar fields, radiation, and hot dark matter as special cases.  

\subsection{Sound Horizon and Scalar Fields}
\label{sec:soundhorizon}

\begin{figure}[hbt]
\centerline{ \epsfxsize = 3.25truein \epsffile{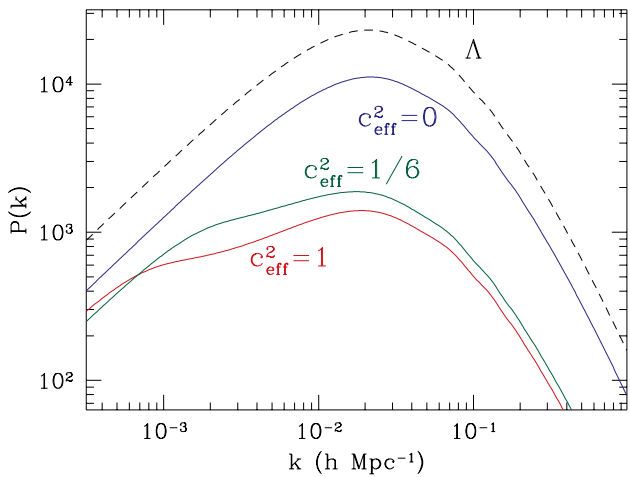}}
\caption{The effective sound horizon and the 
COBE normalized matter power spectrum. 
Raising $c_\eff^2$ from 0 to 1 (solid lines)
introduces a feature between the effective sound horizon at GDM-domination 
and that scale today.
Here $\sigma_8=(0.75,0.29,0.25)$ and $\sigma_{50}/\sigma_8=(0.16,0.17,0.16)$ for 
$c_\eff^2=(0,1/6,1)$.
These models have 
$w_\gdm=-1/6$, $c_\vis^2=0$, 
$\Omega_{\rm tot}=1$, $\Omega_\gdm=0.65$, $\Omega_b h^2=0.0125$, $h=0.7$ and tilt $n=1$.
For comparison, the corresponding $\Lambda$-model
($w_\gdm \rightarrow -1$, same parameters, $\sigma_8=1.1$, 
$\sigma_{50}/\sigma_8=0.16$, dashed lines),
which fits the current large
scale structure data,
is also shown.}
\label{fig:tkceff}
\end{figure}

The introduction of a stabilization scale or effective sound horizon
$s_\eff$ for the GDM places
a feature in the matter power spectrum between that scale
at GDM-domination and today, i.e.
\begin{equation}
s_\eff^{-1}(a=1) < k < s_\eff^{-1}(a=a_g).
\label{eqn:matterfeature}
\end{equation}
The limiting cases are
the scalar-field example, where the effective sound speed is
the speed of light  $c_\eff^2 = 1$ and the pressure-gradient free 
case with $c_\eff^2 = 0$.  In Fig.~\ref{fig:tkceff}, we display the 
effect of varying $c_\eff^2$ on the power spectrum of the matter
holding the background cosmology fixed.
These models have been consistently normalized to the 
COBE CMB anisotropy measurement at large scales
via the fitting form of \cite{Bun97} (1997) (their equations [17]-[20]). 

The amplitude of fluctuations at the fiducial 
$8 h^{-1}$Mpc scale, $\sigma_8$, is affected in three ways.
The presence of a clustering scale reduces the growth rate below it
leading to a relative suppression of small-scale power.
However the absolute amplitude of large-scale fluctuations also changes
with the clustering scale.  As shown in the last section, decreasing
$c_\eff^2$ decreases the amplitude of fluctuations in the matter
relative to the potential fluctuations.  On the other hand, as we shall see in 
the next section, decreasing $c_\eff^2$ also eliminates a source of
CMB anisotropies such that the COBE signal drops relative to the potential
fluctuations.  These two effects compete such that the change in
normalization of $P(k)$ at large scales with $c_\eff^2$ is non-monotonic.  
The shape of the power spectrum, above and below the transition region
of equation~(\ref{eqn:matterfeature}) remains that of a CDM model
with the same $\Omega_m h$ and $\Omega_b h^2$ (c.f.\ $\sigma_{50}/\sigma_8$
in Fig.~\ref{fig:tkceff}).

\begin{figure}[hbt]
\centerline{ \epsfxsize = 3.25truein \epsffile{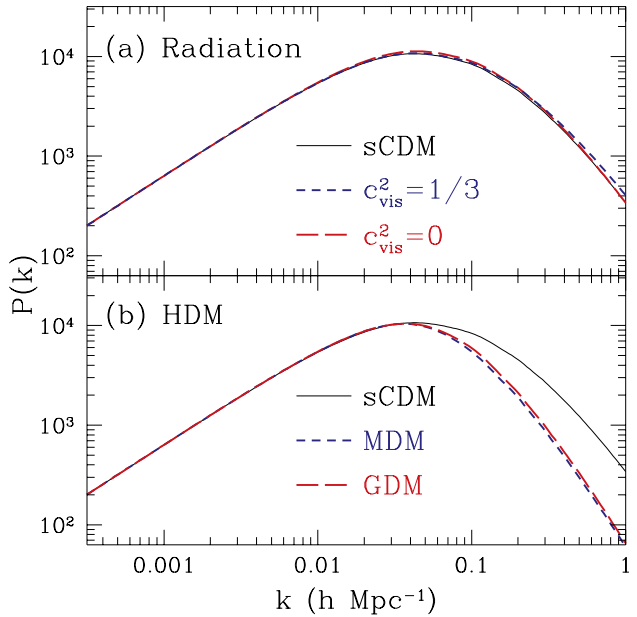}}
\caption{(a) Modeling radiation.
Shown here is the power spectrum for the
model of Fig.~\protect\ref{fig:timepi} where GDM of
$w_\gdm=1/3$ replaces the neutrinos of sCDM.  Altering the viscosity parameter from
$c_\vis^2=1/3$ to $0$ has little effect on the power spectrum although $1/3$ is
a somewhat better approximation at large scales. (b) Modeling HDM.  The features of
the mixed dark matter are well-reproduced by GDM with the same equation of state
and $c_\vis^2=w_\gdm$. The parameters here are sCDM with $\Omega_\nu=0.2$
replacing part of the CDM.}
\label{fig:tkrad}
\end{figure}

\subsection{Viscous Scale and Radiation}
\label{sec:viscousscale}

As discussed in \S \ref{sec:viscosity}, the effect of changing the viscous scale through
$c_\vis^2$ has little effect on the spectrum of matter fluctuations as long
as $|c_\eff^2| > |c_\vis^2|$.  
We show an example in Fig.~\ref{fig:tkrad}a where the neutrinos in sCDM have been 
replaced with GDM as in Fig.~\ref{fig:timepi}.  
In the large-scale structure regime, the difference between these models and sCDM is small and
$c_\vis^2=1/3$ in particular provides an excellent approximation to free-streaming
neutrinos.

\subsection{Time-Dependent Stresses and MDM}
\label{sec:MDM}

In general, the stress parameters ($w_\gdm$, $c_\eff^2$, $c_\vis^2$) may all be 
time-dependent.  An interesting concrete example of such a model is provided by the
mixed dark matter (MDM) scenario where a component of HDM
(e.g. massive neutrinos) is added to the CDM. 
Here the equation of state goes from $w_\hdm = 1/3$ to $0$ as the
neutrinos become non-relativistic.  Fitting to the numerical integration of
the distribution gives
\begin{equation}
w_\hdm  = {1 \over 3} \left[ 1+ (a/a_{\rm nr})^{2 p} \right]^{-1/p},
\end{equation}
with $p=0.872$ and $a_{\rm nr}=6.32 \times 10^{-6} /\Omega_\nu h^2$.  
We can model its behavior as a GDM-component with $w_\gdm = w_\hdm$,
$c_\eff^2=c_\gdm^2$ given by equation~(\ref{eqn:sound}), and $c_\vis^2=w_\hdm$.  
In Fig.~\ref{fig:tkrad}b, we show that this model accurately reproduces
the features of the MDM model as calculated by CMBFAST (\cite{Sel96}\ 1996).
The novelty of this type of model is that the ratio of the
clustering scale to the horizon scale varies, in this case shrinking
with time.   This can have the effect of smoothing out the clustering feature in the 
matter power spectrum (c.f. Figs.~\ref{fig:tkceff} and \ref{fig:tkrad}b).  

An exotic example of this type of model is the self-interacting
dark matter candidate proposed by \cite{Car92}\ (1992), where the equation of
state passes a logarithmically decaying regime
\begin{equation}
w_\gdm = {1 \over 3 \ln (a/\bar a)} \,,
\end{equation}
between the radiation and matter limits
of $w_\gdm=1/3$ and $w_\gdm \propto a^{-2}$. Here $\bar a$ is a constant.

Other examples include a scalar field that rolls from a potential dominated
regime with $w_\gdm=-1$ to a kinetic energy dominated regime with
$w_\gdm=1$ (decaying-$\Lambda$ scenarios, see e.g. \cite{Cob97}\ 1997).  
The scalar field may also go from a rolling to rapidly-oscillating regime
where subhorizon clustering can take place.  Here one must redefine
$c_\eff^2$ to be time-variable as well.
This may occur in certain two-field models where the mass term
can be time-dependent (see Appendix).

\subsection{GDM-Only Models}

The freedom to set the clustering scale well below the horizon ($c_\eff^2 \ll 1$) raises the possibility
that there is only a single component of dark matter with $w_\gdm < 0$,
i.e. CDM is absent.  Conventional scalar field models 
(e.g.\ \cite{Cal98}\ 1998) do not allow this possibility since
perturbations could never grow beyond the small amplitude they possessed at horizon
crossing (but see Appendix). 
%Scalar fields that both oscillate rapidly and roll can potentially 
%behave in this manner (see Appendix).

\begin{figure}[hbt]
\centerline{ \epsfxsize = 3.25truein \epsffile{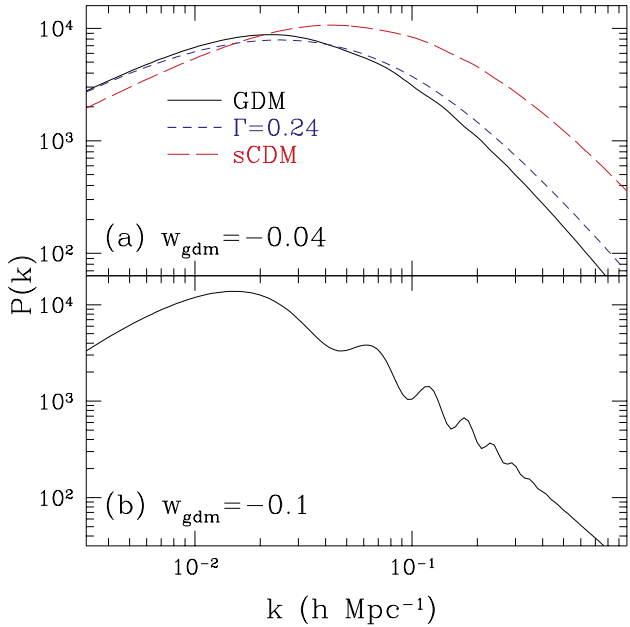}}
\caption{GDM-only Models. (a) $w_\gdm=-0.04$. 
Here an $h=0.65$ model with otherwise sCDM parameters is shown (solid line
$\sigma_8=0.60$, $\sigma_{50}/\sigma_8=0.17$) 
in comparison with sCDM (long-dashed line) and a CDM model with 
$\Gamma=\Omega_m h = 0.24$ (short-dashed line) with a normalization artificially set to match the GDM model.
(b) $w_\gdm=-0.1$.
Acoustic features appear even though the model is a 
critical universe ($\Omega_{\rm tot}=1$, $h=0.65$) 
with big bang nucleosynthesis
baryons ($\Omega_b h^2 = 0.025$).
The suppression of GDM power is counteracted by a strong
blue tilt $n=1.7$
($\sigma_8=0.40, \sigma_{50}/\sigma_8=0.23$).
}
\label{fig:tkonly}
\end{figure}

The lack of a CDM component allows the appearance of 
interesting phenomena in the
matter power spectrum.   The main effect is that the shape parameter of 
CDM is rescaled for a given $\Gamma = \Omega_\gdm h$,
because the relevant scale is the horizon at GDM-radiation equality
given by equation (\ref{eqn:kequality}), i.e.
\begin{equation}
{\Gamma_{\rm eff} \over \Gamma} = 
	\left( { \Omega_r \over \Omega_\gdm} \right)^{
	-3w_\gdm/(1-3w_\gdm)} ,
\end{equation}
where $\Omega_r = 4.17 \times 10^{-5} h^{-2}$ with the usual thermal
history.  Note that $w_\gdm$ appear in the exponent for the 
shape paramter $\Gamma_\eff$ and hence even a mild departure from
zero yields dramatic effects.   In Fig.~\ref{fig:tkonly}a,
we show an example with $w_g=-0.04$ GDM replacing the CDM in a model with
$\Omega_{\rm tot}=1$, $h=0.65$, $\Omega_b h^2=0.0125$, $n=1$ (solid lines).
This model has $\Gamma_{\rm eff}=0.24$ and closely resembles a CDM model with
$\Gamma =\Omega_m h = 0.24$ (short-dashed lines).  Not only does lowering
$w_\gdm$ from zero help the problems of the normalization and shape of the
CDM power spectrum, it also raises the age.

Of course the small change from $w_\gdm=0$ to $-0.04$ in the example above, only
increases the age from to a negligible amount from 10 to 10.4 Gyrs ($h=0.65$).
If we push $w_\gdm$ to $-0.1$, the age is 11.1 Gyrs but $\Gamma_{\rm eff}$ is
too small to account for large scale structure with a scale-invariant $n=1$
spectrum.  This sort of model can be
made viable by blue tilting the initial spectrum.  In Fig.~\ref{fig:tkonly}b, we
show an example with $\Omega_{\rm tot}=1$, $h=0.65$, $\Omega_b h^2=0.025$ and
$n=1.7$ (with a COBE normalization to a model with reionization at $z=65$ in order
that the large tilt be consistent with degree scale anisotropies).  
It is interesting to note that the GDM power is so suppressed that even with this
big bang nucleosynthesis baryon content, acoustic oscillations are visible in a 
critical density model.  In fact, there is an interesting feature in this model
at the $100 h^{-1}$ Mpc scale of $k \sim 0.05-0.07 h $Mpc$^{-1}$.  These models
thus escape the constraints presented in \cite{Eis97} (1997) and may help
to explain the observed $100 h^{-1}$ Mpc excess should it persist.  More generally,
the replacement of CDM with $w_\gdm < 0$ GDM gives one the freedom
to increase the prominence of the acoustic oscillations in the matter power
spectrum.

\section{CMB Anisotropies}
\label{sec:CMB}

The presence of GDM affects the CMB anisotropies both by its
influence on the background expansion (\S \ref{sec:acoustic}) 
and on the gravitational
potential perturbations (\S \ref{sec:ISW}). 
If $w_\gdm<0$ and CDM is also present,
small-angle anisotropies
are affected only by the former since the perturbations
that generated them entered the horizon well when the GDM effects
are negligible.  We discuss the case where it is absent
below in \S \ref{sec:special}.  
Large-angle anisotropies depend mainly on gravitational
potential variations (the ISW effect),
since primary anisotropies have few features that can be
shifted by a change in the background geometry.

\begin{figure}[hbt]
\centerline{ \epsfxsize = 3.25truein \epsffile{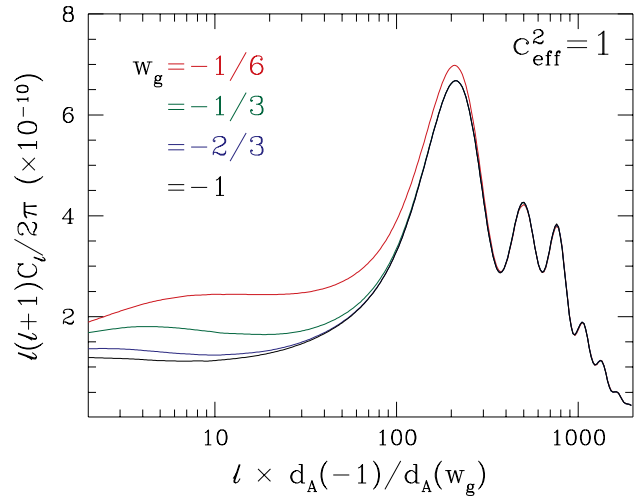}}
\caption{Angular diameter distance, the acoustic peaks and the late ISW effect.
Small-angle anisotropies depend on the equation of state $w_\gdm$ through the
angular diameter distance $d_A$.  Once the angular scale is rescaled to the
fiducial $d_A$ of a $w_\gdm=-1$, $\Lambda$ model, the curves are 
indistinguishable if normalized to small scales. 
Large-angle contributions arise from the ISW effect and are maximal
in these scalar field $c_\eff^2=1$ examples.  Here $\Omega_{\rm tot}=1$, $\Omega_\gdm=0.65$,
$h=0.7$ and $\Omega_bh^2=0.0125$.}
\label{fig:clda}
\end{figure}

\subsection{Acoustic Peaks}
\label{sec:acoustic}

The acoustic peaks in the CMB depend on the photon to baryon ratio, 
the expansion rate at last scattering, and the gravitational
potential, all at last scattering,  as well as the angular diameter distance to last 
scattering (\cite{Acoustic}\ 1996).  
Provided that GDM contributes negligibly to the density at
last scattering, it can only alter the peaks through the last effect.  
Here features shift in
scale with the angular diameter distance, which in a flat universe is 
\begin{equation}
d_A= \eta_0 - \eta_{\rm *} \,.
\end{equation}  
Here 
$\eta_*\equiv \eta(a_*)$ is the conformal time at last scattering
(see \cite{Hu97a} 1997a eq. [22]-[23]).   
We display this effect in Fig.~\ref{fig:clda} where the angular scale
of 4 models with $-1 \le w_\gdm  \le -1/6$ are rescaled via the $d_A$ 
of a fiducial $w_g=-1$ model.
As $w_\gdm$ decreases, $d_A$ increases such that features
are shifted to smaller angular scales.

\begin{figure}[hbt]
\centerline{ \epsfxsize = 3.25truein \epsffile{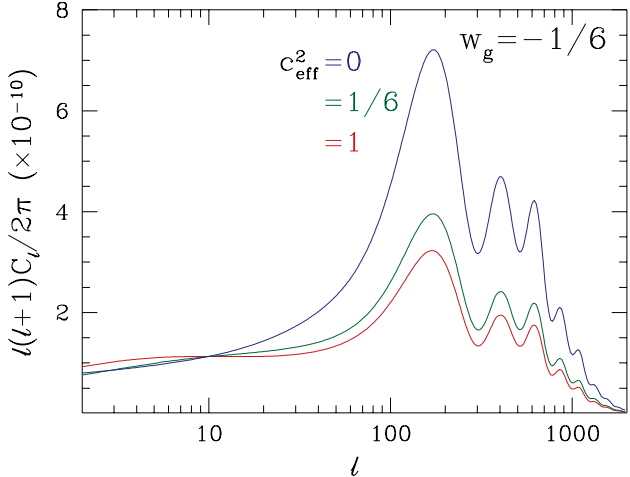}}
\caption{Sound speed effects.  
Decreasing $c_\eff^2$, for
a fixed background cosmology
with $w_\gdm=-1/6$ and the same cosmological
parameters as Fig.~\protect\ref{fig:clda}, 
decreases
the ISW effect at large angles such that COBE normalized models have
lower acoustic peaks.}
\label{fig:clceff}
\end{figure}

\subsection{ISW Effect}
\label{sec:ISW}

A more complicated effect arises from the decay in the gravitational
potential induced by the GDM.  This is called the late ISW effect and
produces a contribution to $C_\ell$ as
\begin{equation}
C_\ell^{(\rm ISW)} = {2 \over \pi} \int {d k \over k} 
k^3 \left[ \int_{\eta_*}^{\eta_0} d\eta (\dot \Psi - \dot \Phi)j_\ell(k(\eta_0-\eta)) \right]^2,
\end{equation}
which appears at large angles.  On 
small scales, the photons can traverse many wavelengths of the
fluctuation during the decay time and thus destroy the 
coherence of the gravitational redshifts.  The cancellation is expressed
by the integral over the oscillatory Bessel function.

\begin{figure}[hbt]
\centerline{ \epsfxsize = 3.25truein \epsffile{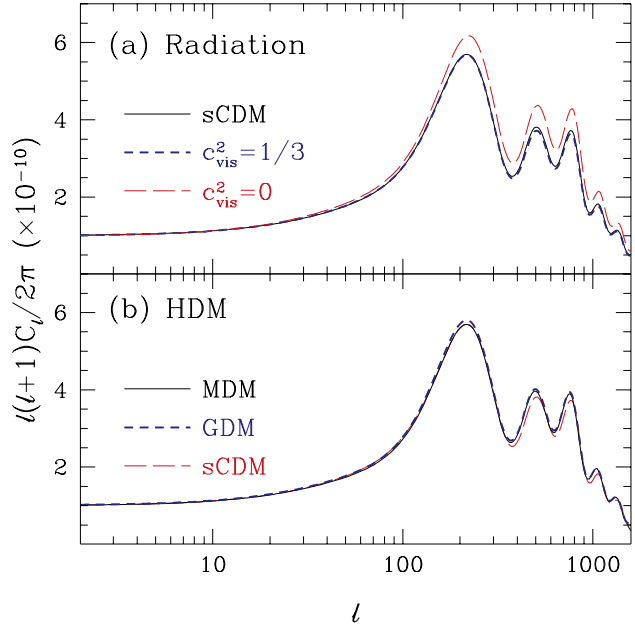}}
\caption{(a) Modeling radiation.
Shown here are the anisotropies for the model where the
neutrinos of sCDM are replaced with
GDM $w_\gdm=1/3$ as in Fig.~\protect\ref{fig:tkrad}a.
Changing the viscosity parameter alters the anisotropies with $c_\vis^2=1/3$ best
approximating the neutrinos.  (b) Modeling HDM (same parameters as
Fig.~\protect\ref{fig:tkrad}b).  The GDM model accurately reproduces the features
of MDM which are themselves only slightly different from sCDM.}
\label{fig:clrad}
\end{figure}

The effect is minimized if $c_{\eff} \ll 1$.  
A concrete example of this is the MDM
scenario where the hot component has $w_\gdm =0$ and an effective 
sound horizon today well below the
particle horizon.   Hence there is essentially no ISW contribution 
in this model (see Fig.~\ref{fig:clrad}).  
More generally, there will be a small effect if $w_\gdm \ne 0$ due 
to the mild potential variation from the change in the equation of
state (see eq.~[\ref{eqn:phiw}]).  

The effect is maximized for $c_{\eff}=1$, 
as is the case for scalar field models shown in Fig.~\ref{fig:clda}.  
Here the potential 
decays due to pressure support of the GDM-fluctuations during GDM-domination.
The crucial aspect is that the decay occurs as soon as the perturbation crosses
the horizon so that the photons have not had sufficient time to cross the perturbation.
The effect thus monotonically decreases as $c_{\eff}^2$ decreases to zero. 
In Fig.~\ref{fig:clceff}, we display this trend in $C_\ell$.  Here, we properly
normalize the spectrum to the COBE detection, which corresponds roughly to $\ell=10$.  
Thus as $c_\eff^2$ increases, the height of the acoustic peaks decreases relative to the
ISW-boosted large-scale anisotropy.  This also has
the consequence of decreasing the normalization of the matter power 
spectrum as we have seen in Fig.~\ref{fig:tkceff}.

\subsection{Special Cases}
\label{sec:special}

It is worthwhile to consider a few special cases to further illustrate the
range of phenomena and show how more conventional candidates are recovered. 
As we have seen in \S \ref{sec:viscosity}, radiation can be modeled 
through the viscous parameter $c_\vis^2$.
Changing $c_\vis^2$ has a greater effect on the
CMB than on the matter power spectrum since it enters directly into 
the evolution of the gravitational potentials (see eq.~[\ref{eqn:Poisson}]).  
In Fig.~\ref{fig:clrad}a, we
show the model of Figs.~\ref{fig:timepi} and \ref{fig:tkrad}a,
where the neutrinos in sCDM are replaced by GDM of $w_\gdm=c_\eff^2=1/3$.
The model with $c_\vis^2=1/3$ yields an excellent approximation to the neutrinos,
whereas that with $c_\vis^2=0$ differs by $\sim 20\%$ from the sCDM results
(see also \cite{Hu95}\ 1995).

Likewise the GDM model for a hot component presented in \S \ref{sec:MDM}
accurately reproduces the anisotropies of an MDM model (see Fig.~\ref{fig:clrad}b).
Note that this model has essentially no late ISW effect and 
differs from sCDM (long-dashed lines) 
only through the small change in the expansion rate and gravitational
potentials due to the presence of the hot component at last scattering.
MDM models have larger small scale anisotropies due potential decay
from the radiation pressure
of the hot component
(\cite{Ma95}\ 1995; \cite{Dod96}\ 1996).
These results in conjunction with the analogous ones for the matter power spectrum
presented in Fig.~\ref{fig:tkrad} demonstrate that the GDM ansatz also allows
one to model the full range of leading particle dark matter candidates.

\begin{figure}[hbt]
\centerline{ \epsfxsize = 3.25truein \epsffile{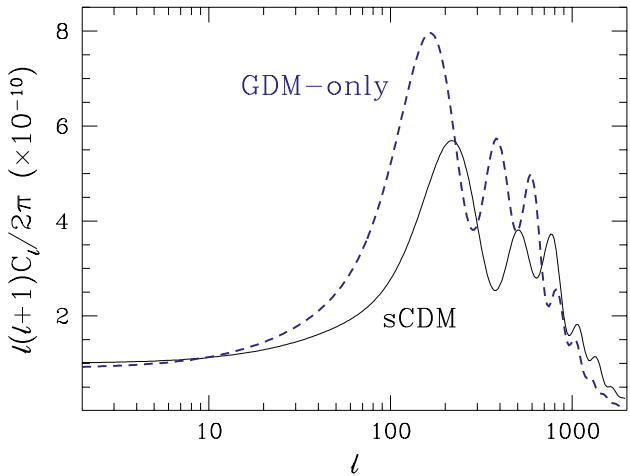}}
\caption{A GDM-only Model. 
CDM is replaced with GDM of $w_\gdm=-0.04$ ($c_\eff^2=c_\vis^2=0$)
as in the model of Fig.~\protect\ref{fig:tkonly}a.  
Compared with the sCDM model, 
the acoustic
peaks are of higher amplitude and larger angle as discussed in the text. }
\label{fig:clxdom}
\end{figure}

Finally, we show the CMB anisotropies for the single dark matter component model
of Fig.~\ref{fig:tkonly}a.  Models of this type tend to have enhanced degree scale
anisotropies.  Because radiation domination is extended, there is more decay 
in the gravitational potentials due to radiation pressure support.  This leads to
early ISW contributions around GDM-radiation equality.  Furthermore, the sound horizon
at last scattering increases due to the extended period of radiation domination.  This
moves the acoustic peaks to slightly larger angles.

\section{Conclusions}
\label{sec:conclusions}

We have introduced a parameterization of a generalized dark matter component
based on three quantities which can vary in time but not in space.  These
are the equation of state $w_\gdm$, the effective sound speed $c_\eff^2$, and
the viscosity parameter $c_\vis^2$. Combinations of these parameters recover
all currently popular candidates for dark matter either exactly or to high
accuracy, e.g. cold dark matter, radiation, hot dark matter, and scalar fields.
Note that the computational costs of modeling the HDM is the same as
CDM, in sharp contrast to a full solution to the energy-dependent
Boltzmann equation
for massive neutrinos (\cite{Ma95}\ 1995).

In this general study, we have shown how the clustering scale of the GDM 
is not in general specified by its equation of state.   The clustering scale
appears as a feature in the matter power spectrum but only a weak enhancement
of CMB anisotropies at large angles.  In fact, the contribution is bounded by 
the scalar field case of $c_\eff^2=1$ for any given equation of state.  There
exist a class of models where GDM clustering has dramatic effects on large
scale structure but no effect on the CMB 
(within cosmic variance).
Conversely, altering the viscosity parameter of the dark matter affects the
CMB anisotropies more strongly than the matter spectrum.  An example of this
behavior is provided by massless neutrinos.  The difference between massless
neutrinos and a perfect fluid with $w_\gdm=1/3$ produces an observable difference
in the CMB anisotropy spectrum.      Finally models exist where
only a single species of dark matter is necessary in contrast to
scalar field and MDM models where a comparable amount of CDM
must also be present to form structure.

While observations currently do not force one to consider the 
the exotic types of dark matter studied here, this situation may soon change
with the high precision measurements expected from the Microwave Anisotropy
Probe (MAP) on
the CMB side and 
the 2 Degree Field (2DF) and Sloan Digital Sky Survey (SDSS) on the
galaxy-survey side (\cite{Hu98b} 1998).  
The freedom to alter large scale structure in relation to the CMB
uncovered here may then be essential in the reconstruction of the cosmological
model.   This is especially true since the ambiguity introduced by the initial
spectrum of fluctuations is removed once the CMB and large scale structure are
measured at the same physical scale.  Here 
we have exposed the aspects of the dark matter to which the CMB and 
large scale structure are and are not sensitive.   These phenomenological aspects,
once observationally determined, should aid isolation of a viable
physical candidate for the dark matter. 

\noindent
Acknowledgements: I thank M. White for allowing me to modify
his Boltzmann code for this work as well as D.J. Eisenstein and D.N. Spergel
for useful discussions.  This work was supported by the W.M. Keck Foundation and
NSF PHY-9513835.

\appendix

\section{Scalar Fields as GDM}

In this appendix, we demonstrate that scalar fields are recovered
exactly as a limiting case of the GDM ansatz and discuss a few 
special cases.  A minimally coupled
scalar field $\varphi$ with the Lagrangian
\begin{equation}
{\cal L} = -{1 \over 2} \sqrt{-g} \left[ g^{\mu \nu} \partial_\mu \varphi
        \partial_\nu \varphi + 2V(\varphi) \right] \,,
\end{equation}
and small perturbations $\varphi = \phi_0 + \phi_1$,
obeys 
\begin{equation}
{\ddot \phi_0} + 2{\dot a \over a} \dot\phi_0 +a^2 \scpot_{,\varphi} = 0\,,
\label{eqn:phi0}
\end{equation}
for the background field and
\begin{eqnarray}
\ddot \phi_1 &=& - 2{\dot a \over a} \dot \phi_1
- (k^2 + a^2 \scpot_{,\varphi\varphi})\phi_1  \nonumber\\
&& +  (\dot h_v - 3\dot h_\delta) \dot\phi_0
	- 2 a^2 \scpot_{,\varphi} h_v \,, 
\label{eqn:phi1}
\end{eqnarray}
for the perturbations.  Recall that 
the metric perturbations $h_v$ and $h_\delta$
were defined for the synchronous and Newtonian gauges in 
equation (\ref{eqn:metric}). 

From the stress-energy tensor
\begin{equation}
T^{\mu}_{\hphantom{\mu}\nu} = \varphi^{;\mu} \varphi_{;\nu}
	- {1 \over 2} (\varphi^{;\alpha} \varphi_{;\alpha} + 2 \scpot) 
	\delta^{\mu}_{\hphantom{\mu}\nu} \,,
\end{equation}
we can associate
\begin{eqnarray}
\rho_\phi &=& { 1 \over 2} a^{-2} \dot \phi_0^2 + \scpot \,,\nonumber\\
p_\phi    &=& { 1 \over 2} a^{-2} \dot \phi_0^2 - \scpot \,,
\end{eqnarray}
for the background and
\begin{eqnarray}
\delta \rho_\phi 
&=& a^{-2}(\dot\phi_0 \dot\phi_1-\dot\phi_0^2 h_v)
+\scpot_{,\varphi} \phi_1 \nonumber\,,\\
\delta p_\phi &=&  \delta \rho_\phi -2\scpot_{,\varphi}\phi_1\,,  \nonumber\\
(\rho_\phi+p_\phi) v_\phi &=& a^{-2} k \dot\phi_0 \phi_1
        \nonumber\,,\\
p_\phi \pi_\phi & = & 0\,,
\label{eqn:scalarfieldpert}
\end{eqnarray}
for the perturbations. 

The equations of motion can now be rewritten as 
\begin{equation}
\dot\rho_\phi = - 3( \rho_\phi + p_\phi){\dot a \over a} \,,
\end{equation}
for the background and
\begin{eqnarray}
(\delta\rho_\phi)\,\dot{\vphantom{p}}
&=& 
	-(\rho_\phi+p_\phi)(k v_\phi + 3\dot h_\delta)
- {\dot a \over a} \Big[ 6 \delta \rho_\phi \nonumber\\
&&
+ 9 {\dot a \over a}
 (1-c_\phi^2)(\rho_\phi + p_\phi)v_\phi/k \Big]\,,\nonumber\\
   \,   [(\rho_\phi + p_\phi) v_\phi/k]\,
\dot{\vphantom{p}}
&=& - (1 + 3 c_\phi^2) (\rho_\phi + p_\phi) {\dot a \over a} v_\phi/k  \nonumber\\
&& + \delta\rho_\phi + (\rho_\phi+ p_\phi) h_v  \,.
\end{eqnarray}
One can rewrite the equations (\ref{eqn:density}), (\ref{eqn:continuity}),
and (\ref{eqn:euler}) for the GDM as
\begin{eqnarray}
(\delta\rho_\gdm)\,\dot{\vphantom{p}}
&=& 
	-(\rho_\gdm+p_\gdm)(k v_\gdm + 3\dot h_\delta)\nonumber\\
&&
- 3 {\dot a \over a} p_\gdm \Gamma_\gdm 
- 3 (1+c_\gdm^2) {\dot a \over a} \delta \rho_\gdm
 \nonumber\\
\,[(\rho_\gdm + p_\gdm) v_\gdm/k]\,\dot{\vphantom{p}}
&=& - 4 {\dot a \over a} (\rho_\gdm+p_\gdm) v_\gdm/k +
   c_g^2 \delta\rho_\gdm 
\nonumber\\
&&
+ p_\gdm\Gamma_\gdm 
	- {2 \over 3}(1-3K/k^2)p_\gdm \pi_\gdm 
	\nonumber\\
&& + (\rho_\phi+ p_\phi)k h_v  \,.
\label{eqn:gdmscalarfields}
\end{eqnarray}
Employing equation (\ref{eqn:gammaansatz}) and (\ref{eqn:pi}),  
we find
that GDM with $w_\gdm = p_\phi/\rho_\phi$, 
$c_\eff^2=1$ and $c_\vis^2=0$ corresponds to a scalar field component.

Although exact, there are a few novel aspects concerning scalar fields
that are hidden in this representation.  The equation of state
$w_\gdm$ encodes the potential $\scpot$ but is {\it not} completely
specified by it.  The equation of state is also dependent on how
the field rolls in the potential as a function of time and this
is dependent on the expansion rate which acts a frictional term in
equation~(\ref{eqn:phi0}).  This fact allows for novel features
such as the attractor solutions investigated by \cite{Fer97}\ (1997).

Furthermore quadratic potentials $\scpot={1\over 2} m^2 \varphi^2$ 
such as found in axion models cause the field and so the
equation of state to rapidly oscillate.  Here equation~(\ref{eqn:phi1})
acts as a driven oscillator such the perturbation does not
stabilize until (\cite{Khl85}\ 1985; \cite{Nam90}\ 1990)
\begin{equation}
k^2 \simgt m \sqrt{G \rho_\gdm} \,,
\end{equation}
despite the fact that $c_\eff^2=1$.    
Although the GDM description of equation~(\ref{eqn:gdmscalarfields})
formally holds, it is impractical to solve because of the the widely-separated 
expansion and oscillation time scales. One can however
model this system by time averaging the oscillations and setting
$w_\gdm=0$ and $c_\eff^2 \ll 1$ as is commonly done for axion models. 

The field may be both rapidly oscillating and slowly rolling 
in certain two-field models (see e.g. \cite{Kod84}\ 1984 for multifield
modifications to the equations of motion).
Since the field evolution depends only on 
potential derivatives, the axion-type instability 
exists on short time scales.
Nevertheless on the expansion time scale, the rolling 
contributes to the background density and pressure.
Models of this type can have a net $w_\gdm<0$ but can cluster 
well below the horizon scale.

\end{document}